**O Espaço em Aristóteles: da bidimensionalidade do *topos*
às seis *diastaseis* que definem os animais**

Francisco Caruso & Roberto Moreira Xavier de Araújo

Centro Brasileiro de Pesquisas Físicas

**1. Introdução**

O conceito de *espaço* ocupa um papel central tanto na Filosofia quanto na Física (CASSIRER, 1923; JAMMER, 1993), cuja evolução tem influenciado mudanças culturais e sociais (CARUSO & MOREIRA, 2017). Por outro lado, uma discussão adequada do significado do espaço na Ciência ou na Filosofia não deve desconsiderar a questão de sua *dimensionalidade*. Isto porque certo fenômeno físico pode depender não apenas do espaço no qual se desenvolve, mas, dado um particular tipo de espaço, pode ainda depender do número de dimensões desse espaço. Essa possibilidade foi muito pouco explorada na História da Física até o século XX, quando esta possível relação foi colocada de forma impecável, desvelando novos horizontes, pelo físico Paul Ehrenfest:

> Qual o papel desempenhado pela tridimensionalidade do espaço nas leis fundamentais da Física? (EHRENFEST, 1920, p. 440).

Dentre tais leis, podem-se incluir as leis do *movimento* e as leis das interações físicas fundamentais.

Desde a década de 1980, os autores têm se ocupado de questões físicas e epistemológicas relacionadas ao conceito de espaço e, em particular, se dedicado a compreender a tridimensionalidade do espaço.

No âmbito da discussão geral sobre o espaço e sua dimensionalidade, a posição de Aristóteles é da maior relevância, como se terá oportunidade de argumentar.

Significativamente, o Estagirita trata diretamente dessa questão em três livros: *Physica*, *De caelo* e *De incessu animalium*. No primeiro, usando a terminologia moderna, ele considerava a dimensionalidade (d) do espaço igual a 2; no segundo, d=3 e no terceiro, d=6.

O fato de Aristóteles ter considerado necessário discutir essa questão em três obras distintas sugere que ele percebeu, no momento mesmo em que estabelece a primeira sistematização da Filosofia, o papel fundamental da questão do *espaço*. Em particular, trata-se de uma importância tão abrangente ao ponto de abarcar a Astronomia, a Física na Terra e, de certa forma, a Biologia (GOTTHELF & LENNOX, 1987).

Entretanto, cabe adiantar que tal zelo não o levou a produzir uma visão sintética (unificada) sobre o espaço, que lhe permitesse iluminar o seguinte ponto: como a dimensionalidade do espaço interfere na descrição do Céu, da Terra e dos seres vivos, ou mesmo a determina. Por quê?

Independente disso, ao colocar a dimensionalidade espacial como elemento essencial dos estudos de um espectro tão grande de fenômenos e atribuir-lhe valores diferentes em contextos diversos, Aristóteles, em certo sentido, antecipa a colocação de Ehrenfest mencionada anteriormente.

Nesse trabalho, o que se pretende é apresentar – embora de forma resumida e ainda incompleta – como o problema da *dimensionalidade* é tratado naquele que é talvez o maior de todos os sistemas – o aristotélico. Em especial, queremos destacar o fato de que a *dimensionalidade* é um conceito importante para a compreensão da obra de Aristóteles, a partir da constatação de que é essencial para a compreensão do *movimento* – tema central desse Congresso – especialmente o *self motion* (GILL & LENNOX, 1994; COOPE, 2015). Enquanto para Platão o mundo das formas, que é o real, é essencialmente estático, eterno e belo, para Aristóteles, é a experiência direta do mundo material que nos conduz ao conhecimento. O mundo real é esse que percebemos em movimento, mutável e fugidio. Portanto, entender as origens do *movimento* (tanto dos corpos inanimados como dos animais) é essencial e, para isso, o conceito de dimensão (διαστάσεις)[1] é fundamental, como pretendemos demonstrar.

## 2. A questão da dimensionalidade na Física Moderna

Do ponto de vista da História da Física, três foram os pensadores que deram contribuições diferenciadas e essenciais ao problema da dimensionalidade do espaço: além de Aristóteles, cuja contribuição será discutida na próxima seção, destacam-se

---

[1] Retomaremos esse ponto mais adiante.

Immanuel Kant e Paul Ehrenfest. As relações entre os critérios metacientíficos utilizados pelos três autores para explicar ou impor limites sobre a dimensionalidade do espaço físico e os sistemas de explicação causal dominantes nos seus correspondentes períodos históricos foram discutidos pelos autores em outro artigo (CARUSO & MOREIRA, 1994) e não serão abordadas aqui. No entanto, alguns breves comentários sobre os trabalhos de Kant e Ehrenfest podem ajudar o leitor a se situar melhor.

Foi durante o florescimento do período de mecanização da visão de Mundo, mais precisamente, em 1747, que o jovem Kant tentou compreender por que o espaço é tridimensional, em seu primeiro escrito (KANT, 1747). Sua proposta inicial era mostrar que a tridimensionalidade do espaço *resultava* da natureza da lei de gravitação de Newton. Ele não teve sucesso nessa tarefa, como foi detalhadamente discutido pelos autores (CARUSO & MOREIRA, 2015). Entretanto, tal conjectura vai se mostrar muito frutífera, pois, pela primeira vez, havia sido aventado que uma lei da Física poderia depender da dimensionalidade do espaço. Esse foi o ponto de partida que levou Ehrenfest a investigar como a descrição quântica do átomo de hidrogênio depende do número de dimensões espaciais, ou ainda como a estabilidade da solução da órbita terrestre no sistema mecânico Terra-Sol, fundamental para a existência dos seres vivos, pode variar com esse número.

De acordo com o escopo do presente trabalho, é relevante dizer que, para enfrentar esse problema, Kant foi obrigado a refletir sobre o modo pelo qual a *matéria corpórea* se relaciona com a *interação das substâncias físicas*. Em outras palavras, como é possível expressar essa interação em termos universais de causa e efeito, e em que modo a *matéria* é capaz de alterar o estado da *alma* por meio de forças que possuem em seu movimento.[2]

Em última análise, sabe-se hoje (CARUSO & MOREIRA, 2015) que Kant, em realidade, propôs uma justificativa para a tridimensionalidade da *extensão* e não do *espaço*. Isto decorre do fato de ele considerar o espaço como algo não perceptivo, como um produto de um esforço intelectual voltado para estabelecer algum tipo de *ordem*, a partir das coisas inteligíveis. Esse *espaço* lhe aparece, portanto, como objeto de estudo da Geometria, e não da Física. O que realmente impressiona a *alma*, o que é perceptível, são os objetos espacialmente extensos: é a matéria que causa efeito sobre outras substâncias.

---

[2] Para o jovem Kant, o estado interno da *alma* pode ser entendido como o *status repraesentativus universi*.

Não obstante seu insucesso em 1747, devemos destacar, como mérito de Kant, o fato de ele ter imaginado a possibilidade de existência de espaços com um número diferente de dimensões, antes que houvesse uma teoria para estes tipos de espaço. Será a descoberta das geometrias não euclidianas (SOMMERVILLE, 1911), no século XIX, que dará impulso a estas questões. Portanto, o problema da dimensionalidade seria também um problema de Matemática.

Quanto a isso, o físico inglês Gerald James Whitrow chama a atenção de que o problema da dimensionalidade do espaço apresenta sempre um caráter dual, envolvendo a Física e a Matemática (WHITROW, 1955, pp. 13-31). Segundo ele, primeiro é necessário que se questione o significado de um espaço ter certo número de dimensões – e isto diz respeito ao domínio da Matemática. Depois, segue-se a questão de *por que* este número é precisamente 3. A este segundo ponto, espera-se que a Física possa contribuir, de forma significativa, elaborando um conhecimento mais profundo das *peculiaridades que distinguem o espaço tridimensional* de outros, postas em destaque na pergunta do matemático alemão Hermann Weyl:

> (...) se Deus, ao criar o Mundo, decidiu fazer o espaço tridimensional, pode-se chegar a uma explicação 'razoável' deste fato, desvelando tais peculiaridades? (WEYL, 1949).

Ora, claro está que, desde a revolução galileana (KOYRÉ, 1966), passa a existir uma notável interdependência entre a Física e a Matemática na descrição da Natureza, passando esta última a ser vista cada vez mais como a linguagem adequada à Física. No entanto, do ponto de vista lógico, a Matemática tem suas limitações intrínsecas, como aquelas impostas pelo Teorema de Gödel (NEWMAN & NAGEL, 1958). Nesse sentido, quando se constroem conceitos fundamentais para qualquer Teoria Física – como o de *espaço físico* – a partir de conceitos matematicamente bem definidos – como o de espaço geométrico –, torna-se extremamente complexo e intricado explicitar os efeitos dessas limitações, inerentes à Matemática, sobre a Física (SCHENBERG, 1985; COSTA & DORIA, 1991; BARROS, 1991); em particular, seus reflexos sobre as qualidades específicas do espaço físico, como a dimensionalidade. *Ipso facto*, a dualidade a que se refere Whitrow parece-nos hoje injustificada do ponto de vista epistemológico, embora esteja na origem histórica da abordagem moderna do problema da dimensionalidade (JAMMER, 1993). Além disto, injustificada, por exemplo, quando se considera a profunda relação entre Geometria e Física, contida no projeto do físico

Albert Einstein de geometrizar a teoria da Gravitação. Como bem observou o historiador e filósofo da ciência israelense Max Jammer:

> Foi Einstein quem esclareceu como a geometria (...) cessa de ser uma ciência axiomático-dedutiva e torna-se uma entre as ciências naturais; a mais velha de todas, na verdade. (JAMMER, 1993, p. 172).

Toda essa dificuldade, que permeia a Física Contemporânea, resume-se em outro comentário de Jammer:

> (...) a estrutura do espaço físico não é, em última análise, nada de dado na natureza ou de independente do pensamento humano. É uma função do nosso esquema conceitual. (JAMMER, 1993, p. 173).

Foi refletindo sobre essas coisas que fomos remetidos à Obra de Aristóteles.

### 3. Da Física Moderna à Filosofia de Aristóteles

Em um grande sistema filosófico – como o aristotélico – deparamo-nos, a todo instante, com questões que tangenciam, explícita ou implicitamente, a conceituação do espaço: o esclarecimento das questões referentes ao problema do espaço pode, muitas vezes, lançar luz sobre o sistema como um todo.

Chamou-nos atenção, em um primeiro momento, como já mencionado, o fato de Aristóteles ter tratado desse problema em vários pontos de sua obra, com conclusões, aparentemente, muito diferentes acerca das *diastaseis*. *Grosso modo*, podemos dizer que é na *Physica* e no *De caelo*, principalmente, que Aristóteles trata da questão do espaço de modo explícito. Já a questão da dimensionalidade do espaço é abordada também, com outro enfoque, nos tratados biológicos, especialmente na Marcha dos Animais.

Segundo o físico e historiador da ciência francês Pierre Duhem, Aristóteles, no *Organon*, concebe o espaço como a soma total de todos os lugares ocupados pelos corpos (DUHEM, Pierre, 1913-17). Na *Física*, no entanto, a rigor, o Estagirita desenvolve apenas uma teoria do "lugar" – do *topos* –, ou seja, uma teoria das posições no espaço (DUHEM, 1985).

O *topos* aristotélico é definido sobre uma estrutura filosófica bem precisa, que se fundamenta no ideal de *Kosmos* e no *horror vacui*. Nela, o Estagirita tenta contemplar a possibilidade do *movimento*[3] e admite o princípio de impenetrabilidade da matéria. A não aceitação do vácuo, por Aristóteles, *per se* determina que sua teoria do lugar se contraponha à atomista e à platônica, ambas incompatíveis com sua Física.

Para Aristóteles, o espaço – ou melhor, o *lugar*[4] – não existe independentemente da matéria,[5] uma vez que ele o define como o limite adjacente ao corpo nele contido (ARISTÓTELES). Esta afirmativa encontra suporte em vários trechos do Livro IV da *Física*, como, por exemplo:

> O lugar é pensado para ser diferente de todos os corpos que vêm para estar nele e se substituem mutuamente (208b, 5).

Em outra passagem do Livro IV, vê-se que, ao contrário da "forma" e da "matéria", o "lugar", enquanto um limite, é separável das coisas:

> (...) O lugar de uma coisa não é nem uma parte nem um estado dela, mas é separável disso. O lugar pode ser pensado como sendo algo como um navio – o navio sendo um lugar transportável. Mas o navio não faz parte da coisa (209b, 25).

Em outra citação, lê-se:

> Assumimos primeiro que o lugar é o que contém aquilo do qual é o lugar, e não faz parte da coisa; novamente, o lugar primário de uma coisa não é nem menor nem maior do que a coisa; de novo, o lugar pode ser deixado para trás e é separável (211a, 1).

Depois de ter eliminado a possibilidade de interpretar o *topos* como "forma", como "matéria" ou como "um tipo de extensão contida entre as extremidades de um corpo", Aristóteles afirma, no Livro IV, que

> o lugar necessariamente é o limite do corpo que contém que está em contato com o corpo contido. E chamo 'corpo contido' aquele que possa ser movido mediante deslocamento (212a, 5).

Em uma linguagem moderna, essa definição de *topos* encontrada na *Física* corresponde a uma superfície bidimensional; para Aristóteles, *tridimensionalidade* é um atributo somente dos *corpos* e não do *espaço*. Tal definição e suas consequências mais amplas merecem uma discussão muito mais detalhada do que seria possível aqui. Vamos, portanto, nos limitar a enfatizar alguns pontos relevantes para nosso propósito.

---

[3] Entendido, aqui, como mudança de posição.
[4] A rigor, na *Física*, Aristóteles emprega somente o termo "lugar" (*topos*).
[5] Tal qual admitirá Einstein mais tarde em sua teoria da Gravitação.

O primeiro é que a doutrina aristotélica – embora não tenha alcançado uma teoria geral do espaço – foi capaz de oferecer uma clara e precisa definição de *lugar*, ao contrário de outras desenvolvidas mais tarde – como a estoica, por exemplo (SAMBURSKY, 1959), que consideraram o espaço amplamente extenso como um conceito mais ou menos intuitivo. É, precisamente, este tipo de concepção vaga que dominará as mais diversas doutrinas medievais (GRANT, 1981). Na verdade, depois de Aristóteles, somente Isaac Newton e Albert Einstein estabeleceram, de forma precisa e operacional, uma clara e objetiva definição de *espaço físico*.[6]

O segundo ponto é que o Estagirita foi obrigado a admitir um espaço não homogêneo, ou seja, a definir um tipo particular de lugar das coisas – o *lugar natural* –, como consequência da sua concepção dinâmica do *Kosmos*, que pressupõe Deus como ato puro e *causa finalis*. Os movimentos em direção aos lugares naturais restaurariam a ordem cósmica pré-determinada e são, portanto, explicados por uma predestinação cosmológico-teológica (um princípio *teleológico*, portanto ligado à ideia de *fim*); desse modo, a teoria aristotélica de lugar depende de Deus.

A impossibilidade de uma quarta dimensão é defendida por Aristóteles em seu *De caelo*. Logo no início do primeiro parágrafo do Livro 1 (intitulado "A perfeição do Universo"), ele afirma que:

> Uma magnitude, se divisível em um modo, é uma linha, se em duas maneiras, uma superfície, e, se em três, um corpo. Além dessas, não há outra magnitude, porque as três dimensões são tudo o que há, e aquilo que é divisível em três direções é divisível em todas (….) (268a, 5).

A seguir, o Estagirita apresenta uma justificativa cosmológica desse número três referindo-se a sua divinização sustentada pelos pitagóricos. De fato, no *De caelo*, ele diz:

> Pois, conforme dizem os pitagóricos, o universo e tudo o que está nele é determinado pelo número três, já que o começo, o meio e o fim dão o número do universo e essas três coisas constituem o número da tríade. E, assim, tendo tomado esse número três da natureza como leis (por assim dizer) dela, fazemos uso adicional do número três na adoração dos deuses (268a 10-15).

---

[6] De certa forma, pode-se afirmar que a história do conceito de espaço consiste nas contribuições de Aristóteles, Newton e Einstein e em um enorme conjunto de comentários a elas.

Séculos mais tarde, o astrônomo grego Cláudio Ptolomeu, em seu livro (perdido) *Da Distância*, publicado por volta de 150 a.C., teria dado uma "prova" da impossibilidade de uma quarta dimensão, baseada no fato evidente de que é impossível desenhar uma quarta linha perpendicular às outras três linhas mutuamente perpendiculares. Na verdade isso não é uma prova, mas apenas reforça a ideia de que não podemos visualizar a quarta dimensão, do que não se pode concluir sobre sua inexistência.[7] No século XX, a questão da impossibilidade da quarta dimensão vai ser formalmente tratada, por exemplo, pelo matemático francês Henri Poincaré (POINCARÉ, 1913).

A identificação entre a tri-dimensionalidade do espaço e Deus será recorrente na História da Ciência e será, por muito tempo, um elo essencial na cultura judaico-cristã (CARUSO & MOREIRA, 2017). O astrônomo Johannes Kepler, por exemplo, teria afirmado que o espaço tem três dimensões por causa da Santíssima Trindade (JUNG, 1992; PAULI, 1955).

Também para Simplício, o *topos*, definido na *Física* de Aristóteles, é concebido essencialmente como uma extensão a *duas dimensões* (SIMPLÍCIO). Ao interpretar o espaço vazio tridimensional como se ele fosse material e ao aceitar a impenetrabilidade de um meio material em outro – por exemplo, um cubo de madeira – aposto em tal vazio, Aristóteles conclui que a interpenetração das dimensões do hipotético espaço vazio com o cubo material é também impossível. Logo a dimensionalidade como atributo de algo imponderável estritamente considerado como corpóreo *não poderia ser três*. Este argumento terá um papel importante na discussão, que vai da Idade Média ao

---

[7] Aqui cabe chamar a atenção para o papel que a visão, o olhar, tem na teoria do conhecimento, desde Platão. Com efeito, a parede da caverna é tudo que nos é dado como ponto de partida da reflexão sobre o mundo. Essa tela bidimensional, espelho da retina, terá, desde então, um papel fundamental na descrição do Real. Isso, entretanto, só ficará claro no Renascimento com a invenção da perspectiva e do emprego da Geometria Euclideana como parte essencial da descrição do mundo físico pela Mecânica, desde Galileu – para quem a natureza está escrita em caracteres matemáticos – até Newton. Tal fato vai caracterizar definitivamente a cultura ocidental e se traduzirá nas expressões do vocabulário usual: "evidência, ponto de vista, visão do mundo, teoria e ideia, que identificam conhecimento e visão" (BULCÃO, 2008). Encerrando o comentário, sabe-se hoje que a visão é descrita pela teoria eletromagnética da luz, a qual, do ponto de vista da Física, pode-se mostrar que só é possível ser matematicamente definida em um espaço-tempo com três dimensões espaciais mais uma temporal (WEYL, 1949).

Séc. XVII, sobre a existência de um espaço separado da matéria[8] (GRANT, 1978; 1981).

De volta à concepção aristotélica de *topos*, vimos que ela é fortemente centrada na realidade ontológica dos corpos. A própria consistência desta definição parece pressupor a *imobilidade* do *topos*, como notou Ioannis Philoponus cerca de mil anos depois (VITELLI, 1888). Se, de acordo com o exemplo clássico, colocássemos uma pedra no fundo de um rio de água corrente, o "envelope" de água constantemente em mudança não seria o "lugar" da pedra, pois, do contrário, a pedra imóvel estaria mudando de lugar, o que seria auto-inconsistente.[9] Estava Aristóteles consciente desse problema? É possível que Aristóteles tenha vislumbrado alguma possível inconsistência em seu sistema ao tratar o problema do movimento dos animais? Em caso afirmativo, como o problema foi contornado? Segundo o próprio Philoponus, inconsistências desse tipo requerem uma nova definição de "lugar" ou de espaço. Seu argumento é que a natureza do espaço deve ser buscada no volume incorpóreo tridimensional cujas extensões são o comprimento, a largura e a profundidade[10] (VITELLI, 1888, p. 567), como havia sido feito por Descartes. Os comentários de Philoponus, Simplício e tantos outros comentadores medievais sugerem que se retorne aos textos de Aristóteles em busca de respostas a essas perguntas.

Ainda segundo a relação entre os conceitos de *movimento* e *topos*, deve-se ponderar que o *lugar* aristotélico não é inerte. De fato, para Aristóteles, o movimento natural das coisas – como fogo, terra e similares, para citar os seus exemplos – mostra não apenas que o lugar é *algo*, mas também que ele exerce certa influência sobre o *movimento*. Enfim, o *movimento natural* é governado pelo princípio teleológico segundo o qual todos os corpos se deslocam para alcançar ou reestabelecer a perfeição, tendo assim a *ordem* (o *kosmos*) como *causa finalis*. Em sua filosofia, ele pretende que o *lugar* exerça também alguma influência sobre as coisas, enquanto agente ordenador, de forma semelhante ao que fez Kant posteriormente, como já mencionado.

---

[8] Em especial, podemos citar René Descartes. Em seus *Princípios*, tentando acomodar a tese de Aristóteles de que, como a interpenetração de dimensões é impossível, não existe espaço separado dos corpos, o filósofo francês identifica "espaço, ou lugar interno" com a "substância corpórea nele contida" (DES CHENES, 1996, pp. 360-1.) Além disso, Descartes sustenta que a "extensão em comprimento, largura e profundidade que constitui o espaço é evidentemente a mesma que constitui o corpo" (*Ibid.*). Esses conceitos cartesianos de extensão e impenetrabilidade são essenciais para seu projeto metafísico de reduzir a matéria à Geometria (POWERS, 1991).
[9] Esse e outros exemplos apontam para o fato de que da definição aristotélica de *topos não* decorrem paradoxos se *não* se considera o movimento.
[10] Raciocínio muito importante para a construção futura de um conceito de espaço absoluto em Newton.

Com relação a esse movimento restaurador da ordem, Aristóteles identifica regiões ou tipos de *lugar*, em número de *seis* [Física, IV, 208b, 12], que propositalmente citaremos em inglês pelo motivo que ficará claro a seguir:

> [Now] these are regions or kinds of place – up [for fire] and down [for earth] and the rest of the six *directions* (208b, 12-3).[11]

É oportuno notar que, ao contrário da versão inglesa supracitada, que emprega o termo *direções*, na maioria das versões em línguas neolatinas, como a francesa,[12] a italiana[13] e a espanhola,[14] traduz-se a palavra grega διαστάσεις[15] por *dimensões*. Apenas na tradução brasileira[16] o termo grego *diastaseis* é traduzido por *direções*. Esse ponto é relevante, pois o termo *dimensão*, pelo menos modernamente, está muito associado à dimensionalidade do espaço. Cabe ainda lembrar que, na edição latina comentada por São Thomas de Aquino, encontra-se o termo *distância*: "Hac autem sunt loci partes et species, sursum et deorsum, et reliquae sex *distantiarum*" (DE AQUINO, 1492).[17] *Distância* ou *separação* é também a tradução portuguesa para *diastasis* dada por Ramiz Galvão em seu *Vocabulário* (GALVÃO, 1994).

A opção pelo termo *direção* parece-nos mais apropriada para exprimir uma qualidade do que potencialmente pode se deslocar de um lugar para outro, *i.e.*, a mobilidade dos corpos. Sendo eles tridimensionais, eles são capazes de se deslocarem ao longo do prolongamento das três direções de sua extensão, em dois sentidos distintos, perfazendo um total de seis possibilidades. Para Remi Brague, o emprego do termo *dimensões* tornou-se tradicional, embora ele sugira efetivamente algo que tenha extensão, ausente no texto de Aristóteles; sendo assim, ele prefere usar a palavra

---

[11] O grifo é nosso.
[12] "(…) mais ce sont là parties et espèces du lieu, je veux dire, le haut, les bas et les autres parmi les six *dimensions*". *Physique (I-IV)*, tome premier, texto estabelecido e traduzido por H. Carteron, segunda edição. Paris: Soc. D'Édition Les Belles Lettres, 1952 (com texto grego a fronte).
[13] "(…) l'alto, e il basso e le altre quattro *dimensioni* sono le parti e le specie del luogo". *Opere*, vol. terzo, *Fisica, Del cielo*, tradução de A. Russo e O. Longo, Roma-Bari, Editori Laterza, 1987.
[14] "(…) estas son las partes y especies del lugar; es decir, el arriba e el abajo, y las restantes *dimensiones* hasta las seis conocidas". *Obras*. Versão do grego, com preâmbulos e notas de F. de P. Samaranch, Madrid, Aguilar, segunda edição, 1973.
[15] O conceito de διαστάσεις foi abordado nesse Congresso por Matheu Damião na palestra intitulada "Os princípios das dimensões no *De caelo* II 2 de Aristóteles. Cf. também sua tese de Mestrado: *As ΔΙΑΣΤΑΣΕΙΣ em Aristóteles: entre as potências da alma e a tridimensionalidade do corpo*.
[16] Aristóteles. O tratado do lugar e do vazio (Física IV, 1-9). Tradução de Arlene Reis, Fernando Coelho e Luiz Felipe Bellintani Ribeiro, a partir da edição do texto grego: *Aristotelis Physica*. Recognovit brevique adnotatione critica instruxit W.D. Rossi. Oxford: Oxonii e Typographeo Clarendoniano, 1992. Publicado em *Anais de Filosofia Clássica*, vol. V (9), pp. 86-105, 2011.
[17] O grifo é nosso.

*distensão* (BRAGUE, 1988) que, como direção, tem uma conotação de algo ligado ao movimento. Além disso, do ponto de vista contemporâneo, após os trabalhos dos matemáticos Georg Cantor e Giuseppe Peano, *dimensão* passa a ser um atributo topológico do espaço. Dada essa acepção moderna do conceito, há ainda que se ter cautela, pois o termo *dimensão* pode sugerir a existência de um expaço externo e independente do corpo, estranha à obra de Aristóteles. Por fim, a escolha do termo *distensão* parece mais apropriada, uma vez que enfatiza a possibilidade de os corpos se deslocarem ao longo do prolongamento natural das três direções de sua própria extensão – cada deslocamento em dois sentidos distintos – perfazendo um total de seis possibilidades. Feitas essas ponderações, de agora em diante procuraremos usar sempre que for o caso, em nosso texto, o termo *diastaseis*.

Vejamos agora como as *distaseis* são definidas por Aristóteles no *De caelo*.

Que a tridimensionalidade é, para Aristóteles, um atributo do corpóreo – suposto completo em magnitude e imutável – se aprende já no primeiro capítulo do seu *De caelo* (Livro I):

> Uma magnitude, se divisível em um modo, é uma linha, se em duas maneiras, uma superfície, e se em três, um corpo. Além dessas, não há outra magnitude, porque as três dimensões são tudo o que há, e aquilo que é divisível em três direções é divisível em todas. (...) uma vez que 'cada' e 'todo' e 'completo' não diferem um do outro com respeito à forma, mas apenas, se de todo modo, em sua matéria e naquilo no qual eles saõ aplicados, somente o corpo entre as magnitudes pode ser completo. Pois apenas ele é determinado pelas três dimensões, ou seja, em um 'todo'. (....) Não podemos passar além do corpo para um outro tipo, como passamos do comprimento para a superfície, e da superfície para o corpo. Pois se pudéssemos, cessaria de ser verdade que o corpo é completa magnitude. Poderíamos passar para além dele somente em virtude de um defeito nele e aquilo que é completo não pode ser defeituoso, uma vez que se extende em todas as direções (268a, 5 e sg.).

Portanto, da leitura da *Física* e do *De Caelo* de Aristóteles, concluímos que o *Ser* – o corpóreo em sua completude – é a *causa materialis* da tridimensionalidade,[18] negada ao *topos*, que é uma extensão bidimensional.

Comentários interessantes sobre a doutrina aristotélica das *diastaseis* em outro contexto (biológico) podem ser encontrados em Brague (*op. cit.*), dos quais destacamos aquele que poderia ter servido de inspiração para a abordagem antrópica do problema da dimensionalidade do espaço (BARROW, 1983; BARROW & TIPLER, 1986), a saber:

> (Aristóteles) diz que os seres vivos são, pela sua natureza, definidos (horizesthai) pelas suas dimensões (diastasis) (BRAGUE, 1988, p. 34).

---
[18] Conclusão semelhante à da Tese de Kant, de 1474.

A passagem à qual Brague se refere aqui é *De incessu* 4, 705ª, 26-28, a qual citamos a seguir em francês:

> Etant donné que les *dimensions*[19] par lesquelles il est naturel que le vivants soint définis sont au nombre de six: le haut et le bas, le devant et le dernière, et ancore la droite et gauche. (BRAGUE, 1988, p. 304).

A ser comparada com a tradução inglesa da mesma passagem dada em *The Revised Oxford Translation*:

> Again, the *boundaries*[20] by which living beings are naturally determined are six in number, superior and inferior, before and behing, right and left. (BARNES, 1985, p. 1098).

Essa diferença de tradução precisa ser entendida ou justificada, principalmente porque, em *De incessu* 2, 704b, 18-21, a edição de Barnes usa o termo *dimensions* e não *boundaries*.

Independente desse detalhe, em nossa opinião, esta é a primeira vez que, de algum modo, se relaciona a *vida* com a *dimensionalidade do espaço* e, mais do que isso, se sugere que a vida, tal qual se conhece, só é compatível com uma certa dimensionalidade espacial, ideia essa sempre recorrente (CARUSO, 2016).

Como já mencionamos, o Estagirita desenvolve uma teoria para as *diastaseis*, baseado no estudo do movimento dos seres vivos na *Marcha dos Animais*. Como sua concepção de espaço é muito diferente daquela de um espaço homogêneo e isotrópico,[21] ele foi levado a considerar a existência de *seis* e não *três diastaseis* (*up-down; forward-*

---

[19] O grifo é nosso.
[20] O grifo é nosso.
[21] Como aquele que será mais tarde sistematizado por Euclides, em seus *Elementos*.

*backward; left-right*). Portanto, ele especula que estas *diastaseis*, que para ele são relacionadas à alma (*psyche*), de alguma forma definem a natureza dos seres vivos.

Eis o último parágrafo do *De incessu*:

> A estrutura dos animais, tanto em suas outras partes, e, em especial, naquelas que dizem respeito à progressão e a qualquer movimento do lugar, é como descrevemos agora. Depois de determinar essas questões, cabe investigar a alma. (19, 714b, 20 e seg.).

**4. Comentários finais**

A título de comentários finais, gostaríamos de enfatizar o quanto a afirmativa de que *o espaço tem seis diastaseis (dimensões)*, por mais estranha que pareça hoje é, na verdade, uma das chaves para a correta compreensão do conceito basilar de *Cosmos* e de movimento em Aristóteles, levando em conta a *vontade*, que diferencia os seres vivos da matéria inanimada.

No Livro IV da Física, quando Aristóteles considera a existência de 6 *diastaseis* para explicar também o movimento de retorno ao lugar natural ele, na verdade, se utiliza de um subterfúgio. Para compreendê-lo, tomemos o exemplo do movimento apenas na vertical. Neste caso, ele considera o que aconteceria com dois elementos primordiais: o fogo (que naturalmente se movimenta para cima) e a terra (cujo movimento natural é para baixo). Assim, o que resulta das duas citações são duas possibilidades para o movimento na vertical. Entretanto, se tivéssemos tratando apenas do movimento de uma pedra retirada de seu lugar natural, cessada a violência que causou o movimento, a pedra tente a cair, nunca a subir. Ao que é inanimado, não lhe resta que, cessada a violência que rompeu a ordem cósmica, retornar ao seu lugar natural. E nesse caso, só há uma *diastasis* relacionada à vertical. Assim, a rigor, na ausência de vontade ou de alma, em um movimento mais geral, as *seis diastaseis*, disponíveis aos seres vivos, se reduzem trivialmente a *três*.

A tridimensionalidade é sempre um atributo do corpóreo e não do espaço ou do *topos*; esse é bidimensional, embora, como recordamos, tal definição de lugar não está livre de paradoxos.

**Agradecimentos**

Optamos por registrar aqui os agradecimentos feitos em primeira pessoa por Francisco Caruso ao apresentar este trabalho no Congresso OUSIA: "Fiz a minha iniciação científica e meu mestrado em Física de Partículas com Alberto Santoro. A partir da amizade que nasceu nesse período, acabei tendo o prazer de conhecer o maestro Cláudio Santoro durante meu doutorado na Itália. Já de volta ao Brasil, em 1996, integrei o grupo (com Flora Simonetti Coelho, Mirian de Carvalho e Roberto Moreira Xavier) que criou a revista *Dialoghi*, da qual fui editor enquanto ela existiu. Para sua seção "Paesaggi e Profili", convidei, por sugestão de Alberto, Fernando Santoro para escrever um artigo em homenagem a seu tio Cláudio. O resultado foi um belíssimo texto e o início de minha amizade com o mais jovem dos Santoro que conheço. Assim como seus dois tios, uma das características mais marcantes do Fernando é sua generosidade. E ele a exerceu mais uma vez conosco, dando a oportunidade e o privilégio a esses dois autores físicos, declaradamente não especialistas em Aristóteles, de poderem falar aqui para uma plateia de estudiosos. Além disso, permitiu que trouxéssemos algumas ideias ainda inacabadas. Por tudo isso o nosso sincero agradecimento a ele". Por outro lado, gostaríamos ainda de agradecer a Matheus Damião, que tivemos o prazer de conhecer no Congresso, por seu interesse em nosso trabalho e pela frutífera troca de ideias.
**Referências bibliográficas**

ARISTÓTELES. Usaremos neste ensaio edição a cura de BARNES, J. *The Complete Works of Aristotle* (the revised Oxford translation), vols. I e II, Princeton: Princeton University Press, 2ª impressão, 1985, com traduções dos próprios autores, a menos que se diga o contrário.

ATMANANSPACHER, Harald. "The hidden side of Wofgang Pauli: An Eminent Physicist's Extraordinary Encounter with Deep Psycology". *Journal of Counsciousness Studies*, v. 3 (2): pp. 112-126, 1996.

BARROS, José Acácio de. *Dois exemplos de indecidibilidade e incompletude em Física*, Tese de Doutorado. Rio de Janeiro: CBPF, 1991.
14

BARROW, J.D. "Dimensionality", *Philosophical Transactions of the Royal Society of London*, v. A310: p. 337, 1983.

BARROW, John D. & & TIPLER, Frank J. *The Anthropic Cosmological Principle*. Oxford: Oxford University Press, 1986.

BRAGUE, Remi. *Aristote et la question du monde*, Paris: Presses Univ. France, 1988.

BULCÃO, Marly. "A noção de imaginação: Bachelard crítico de Sartre", *in* Constança Marcondes CÉSAR e Marly BULCÃO (Organizadoras). *Sartre e seus contemporâneos*. Aparecida: Ed. Ideias e Letras, 2008.

CARUSO, Francisco. "Life and space dimensionality: a brief review of old and new entangled arguments", *Journal of Astrobiology & Outreach*, v. 4 (2): a.n. 152, 2016.

CARUSO, Francisco; MOREIRA XAVIER, Roberto. "*Causa Efficiens versus Causa Formalis:* origens da discussão moderna sobre a dimensionalidade do espaço", *Cadernos de História e Filosofia da Ciência*, Série 3, v. 4 (2): p. 43-64, julho-dezembro de 1994.

CARUSO, Francisco; MOREIRA XAVIER, Roberto. "On Kant's First Insight into the Problem of Space Dimensionality and its Physical Foundations", *Kant-Studien*, v. 106 (4): pp. 547–560, 2015.

CARUSO, Francisco; MOREIRA XAVIER, Roberto. *O livro, o espaço e a natureza: ensaio sobre a leitura do mundo, as mutações da cultura e do sujeito*. São Paulo: Livraria da Física, 2017.

CASSIRER, Ernest. *Philosophie der symbolischen Formen*. Berlin: Bruno Cassirer, 1923. Tradução para o português: *A Filosofia das Formas Simbólicas*. São Paulo: Martins Fontes, vols. 1 e 2, 2004 e vol. 3, 2011.

COOPE, Ursula. "Self-motion as other-motion in Aristotle's Physics", in LEUNISSEN, Mariska (Ed.). *Aristotle's Physics: A Critical Guide*. Cambridge: Cambridge University Press, 2015.

COSTA, Newton Carneiro Affonso da & DORIA, Francisco Antônio. "Undecidability and incompleteness in Classical Mechanics", *International Journal of Theorethical Physics*, v. 30: p. 1041 (1991).

DE AQUINO, Thomas. *In Octo Libros Physicorum Aristotelis Expositio*, 1492. Edição utilizada Taurini: Marietti, 1965, p. 201, IV 282 (4), 8-11.

DES CHENE, Dennis. *Physiologia: Natural Philosophy in Late Aristotelian and Cartesian Thought*. Ithaca and London: Cornell University Press, 1996.

DUHEM Pierre. *Le Système du Monde*. Paris: Librairie Scientifique Hermann, v. 1, p. 197, 1913-1917.
________. *Medieval Cosmology: Theories of Infinity, Place, Void, and the Plurality of Worlds*. Chicago and London: The University of Chicago Press, 1985.

EHRENFEST, Paul. "Welche Rolle spielt die Dreidimensionalität des Raumes in den Grundgesetzen der Physik?", *Annalen der Physik*, v. 61: p. 440, 1920. *Cf.* também seu "In what way does it become manifest in the fundamental laws of physics that space has three dimensions?", *Proc. Amsterdam Acad.*, v. 20: p. 200, 1917 (Reimpresso em KLEIN, M.J. (ed.) *Paul Ehrenfest – Collected Scientific Papers*. Amsterdam: North Holland Publ. Co., 1959, pp. 400-409).

GALVÃO, Romiz. *Vocabulário Etimológico, Ortográfico e Prosódico das Palavras Portuguêsas Derivadas da Língua Grega*. Belo Horizonte: Livraria Garnier, 1994.

GILL, Mary Louise & LENNOX, James G. (Eds.). *Self Motion: From Aristotle to Newton*. Princeton, New Jersey: Princeton University Press, 1994.

GRANT, Edward. "The Principle of the Impenetrability of Bodies in the History of Concepts of Separate Space from the Middle Ages to the Seventeenth Century", *Isis*, v. 69: pp. 551-71, 1978.

\_\_\_\_\_\_\_\_. *Much Ado About Nothing: Theories of Space and Vacuum from the Middle Ages to the Scientific Revolution*. Cambridge: Cambridge University Press, 1981. Veja também Max Jammer, *op. cit.*

KANT, Imanuel. *Gedanken von der wahren Schätzung der lebendigen Kräfte und Beurtheilung der Beweise, deren sich Herr von Leibniz und andere Mechaniker in dieser Streitsache bedient haben, nebst einigen vorhergehenden Betrachtungen, welche die Kraft der Körper überhaupt betreffen*, 1747. Akademie-Ausgabe (AA): GSK, AA 01.

GOTTHELF, Allan & LENNOX, James G. (Editors). *Philosophical Issues in Aristotle's biology*. Cambridge: Cambridge University Press, 1987.

JAMMER, M. *Concepts of Space: the History of Theories of Space in Physics*. Cambridge: Harvard University Press, 3ª edição, 1993.

JUNG, Carl Gustav. *O Homem e seus Símbolos*. Rio de Janeiro: Nova Fronteira, 11ª edição, 1992, p. 307.

KOYRÉ, Alexandre. *Etudes Galiléannes*. Paris: Hermann, 1966.

NEWMAN, James R. & NAGEL, Ernest. *Godel's Proof*. NY: New York University Press, 1958.

PAULI, Wolfgang. "The Influence of Archetypal Ideas on the Scientific Theories of Kepler", *in* JUNG, C.G. & PAULI, W. *The Interpretation of Nature and Psyche*. London: Routledge & Kegan Paul, 1955. Reprinted by Ishi Press, 2012. A esse respeito, veja também (ATMANANSPACHER, 1996).

POINCARÉ, Henri. *Dernières Pensées*. Paris: Ernest Flammarion, 1913.

POWERS, Jonathan. *Philosophy and the New Physics*. London and New York: Routledge, 1991.

SAMBURSKY, S. *Physics of the Stoics*. London: Routledge & Kegan Paul, 1959.

SCHENBERG, Mario. *Pensando a Física*. São Paulo: Ed. Brasiliense, 2ª Ed, 1985.

SIMPLÍCIO. *Física*, 601, *apud* Jammer, *op. cit.*

SOMMERVILLE, Duncan M.Y. *Bibliography of Non-Euclidean Geometry*. Scotland: The University of St. Andrews, 1911.

VITELLI, H. (Ed.). *Ioannis Philoponi in Aristotelis libros quinque posteriores commentaria*. Berlin, 1888.

WEYL, Hermann. *Philosophy of Mathematics and Natural Science*. Princeton: Princeton University Press, 1949.

WHITROW, G.J. "Why Physical Space has three Dimensions", *British Journal for the Philosophy of Science* v. 6: pp. 13-31, 1955.